\documentclass[aip,jcp,reprint,noshowkeys,superscriptaddress]{revtex4-1}
\usepackage{graphicx,dcolumn,bm,xcolor,multirow,amscd,amsmath,amssymb,amsfonts,physics,longtable,wrapfig,bbold,xspace,microtype,siunitx}
\usepackage[version=4]{mhchem}

\usepackage[utf8]{inputenc}
\usepackage[T1]{fontenc}

\usepackage{soul} %I added this to strike through some sentence.

\usepackage{hyperref}
\hypersetup{
    colorlinks,
    linkcolor={red!50!black},
    citecolor={red!70!black},
    urlcolor={red!80!black}
}

\usepackage{listings}
\definecolor{codegreen}{rgb}{0.58,0.4,0.2}
\definecolor{codegray}{rgb}{0.5,0.5,0.5}
\definecolor{codepurple}{rgb}{0.25,0.35,0.55}
\definecolor{codeblue}{rgb}{0.30,0.60,0.8}
\definecolor{backcolour}{rgb}{0.98,0.98,0.98}
\definecolor{mygray}{rgb}{0.5,0.5,0.5}

\definecolor{sqred}{rgb}{0.85,0.1,0.1}
\definecolor{sqgreen}{rgb}{0.25,0.65,0.15}
\definecolor{sqorange}{rgb}{0.90,0.50,0.15}
\definecolor{sqblue}{rgb}{0.10,0.3,0.60}

\lstdefinestyle{mystyle}{
    backgroundcolor=\color{backcolour},
    commentstyle=\color{codegreen},
    keywordstyle=\color{codeblue},
    numberstyle=\tiny\color{codegray},
    stringstyle=\color{codepurple},
    basicstyle=\ttfamily\footnotesize,
    breakatwhitespace=false,
    breaklines=true,
    captionpos=b,
    keepspaces=true,
    numbers=left,
    numbersep=5pt,
    numberstyle=\ttfamily\tiny\color{mygray},
    showspaces=false,
    showstringspaces=false,
    showtabs=false,
    tabsize=2
  }

  \newcolumntype{d}{D{.}{.}{-1}}

\newcommand{\SupMat}{\textcolor{blue}{supplementary material}\xspace}
\newcommand{\tabc}[1]{\multicolumn{1}{c}{#1}}

\newcommand{\mc}{\multicolumn}

\newcommand{\e}{\epsilon}

\newcommand{\IP}{\text{IP}}
\newcommand{\EA}{\text{EA}}

\newcommand{\HOMO}{\text{HOMO}}
\newcommand{\LUMO}{\text{LUMO}}
\newcommand{\HF}{\text{HF}}

\newcommand{\br}[1]{{\mathbf{r}_{#1}}}
\newcommand{\bx}[1]{{\mathbf{x}_{#1}}}

\newcommand{\hS}{\Hat{S}}

\newcommand{\LCPQ}{Laboratoire de Chimie et Physique Quantiques (UMR 5626), Universit\'e de Toulouse, CNRS, UPS, France}
\newcommand{\NEEL}{Universit\'e Grenoble Alpes, CNRS, Institut NEEL, F-38042 Grenoble, France}
\newcommand{\UCAM}{Yusuf Hamied Department of Chemistry, University of Cambridge, Lensfield Road, Cambridge, CB2 1EW, U.K.}

\begin{document}	

\title{Can $GW$ Handle Multireference Systems?}

\author{Abdallah \surname{Ammar}}
	\email{aammar@irsamc.ups-tlse.fr}
	\affiliation{\LCPQ}
\author{Antoine \surname{Marie}}
	\email{amarie@irsamc.ups-tlse.fr}
	\affiliation{\LCPQ}
\author{Mauricio \surname{Rodr\'iguez-Mayorga}}
	\email{marm3.14@gmail.com}
	\affiliation{\NEEL}
\author{Hugh~G.~A.~\surname{Burton}}
	\email{hgaburton@gmail.com}
	\affiliation{\UCAM}
\author{Pierre-Fran\c{c}ois \surname{Loos}}
	\email{loos@irsamc.ups-tlse.fr}
	\affiliation{\LCPQ}

\begin{abstract}
Due to the infinite summation of bubble diagrams, the $GW$ approximation of Green's function perturbation theory has proven particularly effective in the weak correlation regime, where this family of Feynman diagrams is important. However, the performance of $GW$ in multireference molecular systems, characterized by strong electron correlation, remains relatively unexplored. In the present study, we investigate the ability of $GW$ to handle closed-shell multireference systems in their singlet ground state by examining four paradigmatic scenarios.  Firstly, we analyze a prototypical example of a chemical reaction involving strong correlation: the potential energy curve of \ce{BeH2} during the insertion of a beryllium atom into a hydrogen molecule. Secondly, we compute the electron detachment and attachment energies of a set of molecules that exhibit a variable degree of multireference character at their respective equilibrium geometries: \ce{LiF}, \ce{BeO}, \ce{BN}, \ce{C2}, \ce{B2}, and \ce{O3}. Thirdly, we consider a \ce{H6} cluster with a triangular arrangement, which features a notable degree of spin frustration. Finally, the dissociation curve of the \ce{HF} molecule is studied as an example of single bond breaking. These investigations highlight a nuanced perspective on the performance of $GW$ for strong correlation, depending on the level of self-consistency, the choice of initial guess, and the presence of spin-symmetry breaking at the Hartree-Fock level.
\end{abstract}

\maketitle

%%%%%%%%%%%%%%%%%%%%%%%%%%
\section{Introduction}
%%%%%%%%%%%%%%%%%%%%%%%%%

The $GW$ approximation of many-body perturbation theory, as proposed by Hedin, \cite{Hedin_1965} can be regarded as the workhorse of Green's function methods, \cite{MartinBook} as the vast majority of contemporary calculations performed within this theoretical framework are conducted using the $GW$ approximation. The importance of $GW$ is evident in the solid-state community \cite{Aryasetiawan_1998,Onida_2002,Reining_2017} and its influence is now extending to quantum chemistry, where $GW$ has experienced a substantial surge in popularity over the past decade. \cite{Golze_2019,Marie_2023b} 

This widespread adoption can, in part, be ascribed to the emergence of electronic structure packages that provide efficient implementations of the $GW$ equations, \cite{Blase_2011,Deslippe_2012,Blase_2018,Duchemin_2020,Duchemin_2021,Bruneval_2016,vanSetten_2013,Kaplan_2015,Kaplan_2016,Krause_2017,Forster_2022b,Forster_2022a,Forster_2021,Forster_2020,Caruso_2012,Caruso_2013,Caruso_2013a,Caruso_2013b,Iskakov_2019,Sun_2020,Scott_2023} enabling calculations on large-scale molecular systems. \cite{vanSetten_2013,Neuhauser_2013,Govoni_2015,Liu_2016,Vlcek_2017,Wilhelm_2018,Duchemin_2019,DelBen_2019,Forster_2020,Duchemin_2020,Duchemin_2021,Forster_2021,Panades-Barrueta_2023} These software packages have been bolstered by the creation of well-curated and accurate reference values, such as the $GW100$ dataset of van Setten and collaborators \cite{vanSetten_2015} which reports ionization potentials (IPs) and electron affinities (EAs) for 100 small- and medium-sized closed-shell molecules containing a variety of elements and chemical bonds (see also Refs.~\onlinecite{Caruso_2016,Govoni_2018,Maggio_2017,Colonna_2019}). Similar arguments \cite{Rohlfing_1999a,Horst_1999,Puschnig_2002,Tiago_2003,Rocca_2010,Boulanger_2014,Jacquemin_2015a,Bruneval_2015,Jacquemin_2015b,Hirose_2015,Jacquemin_2017a,Jacquemin_2017b,Rangel_2017,Krause_2017,Gui_2018,Blase_2018,Liu_2020,Blase_2020,Holzer_2018a,Holzer_2018b,Loos_2020e,Loos_2021,Monino_2021,McKeon_2022,Monino_2023} can be put forward for formalisms based on the Bethe-Salpeter equations. \cite{Salpeter_1951,Strinati_1988,Blase_2018,Blase_2020} 

$GW$ is often hailed as ``miraculously'' accurate for weakly correlated systems, given its quite reasonable computation cost. \cite{Bruneval_2021} However, it is usually considered inadequate for strongly correlated materials. \cite{Verdozzi_1995,DiSabatino_2016,Tomczak_2017,Dvorak_2019a,Dvorak_2019b,DiSabatino_2022,DiSabatino_2023,Orlando_2023b}
This perception arises because the $GW$ self-energy is contructed based on a polarizability computed as an infinite summation of a specific class of diagrams, known as bubble diagrams, \cite{Gell-Mann_1957,MattuckBook} which are recognized to be relevant primarily in the weakly correlated regime. \cite{Bohm_1951,Pines_1952,Bohm_1953,Nozieres_1958,Scuseria_2008,Lange_2018,Tolle_2023}
In this context, the term ``strong correlation'' denotes a specific form of electron correlation observed, for example, in transition metal oxides (such as Mott insulators \cite{Mott_1949,Imada_1998}), the large-$U$ limit of the Hubbard model, \cite{Hubbard_1963,Lieb_1968,Montorsi_1992} or the low-density regime of the uniform electron gas \cite{VignaleBook,Loos_2016} (where Wigner crystals are formed \cite{Wigner_1934}). Therefore, the assessment of $GW$ and the definition of strong correlation are specifically rooted in strongly correlated systems pertinent to the condensed matter community. The objective of this work is to evaluate whether this assessment stands for strongly correlated systems encountered in quantum chemistry.

Before introducing the systems that we studied to address this question, let us mention that alternative approximations based on Green's functions do exist and have been studied in the strong correlation regime. The $T$-matrix \cite{Baym_1961,Baym_1962,Danielewicz_1984a,Danielewicz_1984b,Barbieri_2007,DickhoffBook,Romaniello_2012,Zhang_2017,Li_2021b,Loos_2022,Zhang_2017,Li_2021b,Li_2023,Orlando_2023a,Orlando_2023b} approximation is based on an alternative infinite summation to the one used to build the $GW$ polarizability, namely a summation of ladder diagrams. \cite{Peng_2013,Scuseria_2013,Berkelbach_2018,Orlando_2023b} This resummation is justified in the low-density limit of the uniform electron gas with short-range interactions. \cite{MattuckBook} It is not clear whether this resummation is adapted to single- or multi-reference molecular systems, where the long-range Coulomb interaction is ubiquitous. Numerous groups have proposed strategies to go beyond the $GW$ approximation, but these have their own theoretical and practical challenges. \cite{DelSol_1994,Shirley_1996,Schindlmayr_1998,Morris_2007,Shishkin_2007b,Romaniello_2009a,Romaniello_2012,Gruneis_2014,Hung_2017,Maggio_2017b,Cunningham_2018,Vlcek_2019,Lewis_2019a,Pavlyukh_2020,Wang_2021,Bruneval_2021,Mejuto-Zaera_2022,Wang_2022,Forster_2022b}

In the present context, strong correlation specifically refers to molecular systems where multiple electronic configurations are nearly degenerate, thus strong correlation is synonymous with static correlation.
A system is considered strongly correlated if there is more than one electronic configuration with a significant weight in the configuration interaction (CI) expansion. Crucially, this definition is contingent on the choice of the underlying orbitals used to construct these electronic configurations, which further blurs the demarcation between weak and strong correlation.
Hence, it is not surprising that several diagnostics have been developed to measure multireference character in different contexts. \cite{Shannon_1948,Nielsen_1999,Ivanov_2005,Fogueri_2012,Ramos-Cordoba_2016,Bartlett_2020,Xu_2024}

We explore the capability of the $GW$ approximation to handle such closed-shell multireference systems in their singlet ground state. To assess this, we examine four distinct quantum chemistry scenarios involving such systems. Firstly, we analyze the potential energy curve of \ce{BeH2} during the insertion of a beryllium atom into a hydrogen molecule, resulting in the linear \ce{BeH2} molecule. \cite{Purvis_1983} This system serves as a prototypical example of strong correlation and has been extensively studied by various authors in different contexts. \cite{Gdanitz_1988,Mahapatra_1998,Mahapatra_1999,Sharp_2000,Kallay_2002,Pittner_2004,Ruttink_2005,Lyakh_2006,Yanai_2006,Evangelista_2011a,Evangelista_2011b,Evangelista_2012} Secondly, we compute the properties of a set of molecules exhibiting, at their respective equilibrium geometries, a variable degree of multireference character. Thirdly, we investigate the \ce{H6} system arranged in a triangular configuration, a system showing a significant amount of spin frustration. Finally, the evolution of the principal IP of the \ce{HF} molecule is studied during its dissociation, which is a stringent test due to the varying amount of dynamical and static correlations as a function of the bond length. \cite{Bauschlicher_1986,Peterson_1995,Ghose_1995,Li_1998,Krylov_1998,Ivanov_2006,EngelsPutzka_2009,Das_2010,Evangelista_2011b} Our key findings reveal a nuanced perspective on the capabilities of $GW$ in describing multireference systems, indicating that it does possess a certain ability to capture their complex electronic structure. Unless otherwise stated, atomic units are used.

%%%%%%%%%%%%%%%%%%%%%%%%%%
\section{A Primer on $GW$}
%%%%%%%%%%%%%%%%%%%%%%%%%%

Here we report the set of equations required to understand and apply the $GW$ formalism and refer the interested reader to dedicated reviews \cite{Aryasetiawan_1998,Onida_2002,Reining_2017,Golze_2019,Marie_2023b} and books \cite{MartinBook,CsanakBook,FetterBook,DickhoffBook} for additional information.

In the four-point formalism, \cite{Starke_2012,Maggio_2017b,Orlando_2023b} the instantaneous Coulomb potential is defined as 
\begin{equation}
	v(12;1'2') = \delta(11') \frac{\delta(t_1-t_2)}{\abs{\br{1} - \br{2}}} \delta(22')
\end{equation}
Here, $\delta(11')$ is the Dirac function, and the integers, e.g.~$1$, serve as shorthand notations for time ($t_1$) and 
spin-space $\bx{}_1 = (\sigma_1,\br{}_1)$ variables for each particle.

In practice, within the $GW$ approximation, we initiate the process by considering a reference 
propagator $G_0$, typically derived from a mean-field model. Therefore, we directly present the coupled integro-differential 
equations governing the $GW$ formalism for $G_{0} = G_{\text{HF}}$. 
The total self-energy is represented as a sum of the Hartree (H), exchange (x), and correlation (c) self-energies, such that 
\begin{equation}
	\Sigma(11') = \Sigma_{\text{H}}(11') + \Sigma_{\text{x}}(11') + \Sigma_{\text{c}}(11')
\end{equation}

The exchange-correlation part, $\Sigma_{\text{xc}} = \Sigma_{\text{x}} + \Sigma_{\text{c}}$, is expressed as a convolution of the interacting Green's function $G$ and 
the dynamically screened Coulomb interaction $W$,
\begin{equation}
	\Sigma_{\text{xc}}(11') = \imath \int d(22')  G(22')  W(12';21')
\end{equation}
where $W$ is determined by the irreducible polarizability $\tilde{L}$, as follows:
\begin{multline}
	W(12;1'2') 
	= v(1 2^{-};1' 2') \\
	- \imath \int d(343'4')  W(14;1'4')  \tilde{L}(3'4';3^+4)  v(2 3;2' 3')
\end{multline}
A sign over an integer, denoted as, for example, $1^{\pm}$, indicates an infinitesimal time shift $t_{1^{\pm}}=t_1 \pm \eta$.
In the $GW$ framework, the irreducible polarizability is approximated by the product of two Green’s functions
\begin{equation}
	\tilde{L}(12;1'2') = G(1 2')  G(2 1')
\end{equation}
while the one-body Green's function $G$ is obtained through a Dyson equation
\begin{equation}
	G(11') = G_{\text{HF}}(11') + \int d(22') G_{\text{HF}}(12) \Sigma_{\text{c}}(22') G(2'1')
	\label{eq:G_Dyson}
\end{equation}

In practical applications, it is advantageous to introduce a set of real-valued spin-orbital basis functions, denoted as $\{\varphi_p\}$, with corresponding energies $\{\epsilon_p\}$ for describing quasiparticles. This approach enables us to write down the Lehmann representation of $G$ in the following manner
\begin{equation}
	G(\bx{1} \bx{1'};\omega) 
	= 
	\sum_i \frac{\varphi_i(\bx{1}) \varphi_i(\bx{1'})}{\omega - \epsilon_i - \imath \eta} + 
	\sum_a \frac{\varphi_a(\bx{1}) \varphi_a(\bx{1'})}{\omega - \epsilon_a + \imath \eta}
	\label{eq:G_Lehman_quasipart}
\end{equation}
Here, we follow the common practice of using $a, b, \dots$ to represent states above the Fermi level (virtual orbitals) and $i, j, \dots$ for states below (occupied orbitals). The indices $p, q, \dots$ denote arbitrary (i.e., occupied or virtual) orbitals. The calculation of quasiparticle energies and their corresponding Dyson orbitals will be the focus of the upcoming discussion.

Performing various Fourier transforms and projecting onto the spinorbital basis enable us to derive the analytical expression of the matrix elements associated with the correlation part of the self-energy
\begin{equation}
	\left[\Sigma_{\text{c}}(\omega)\right]_{pq} =
	\sum_{mi} \frac{M_{p i}^{m} M_{q i}^{m}}{\omega + \Omega_m - \epsilon_i - \imath \eta} +
	\sum_{ma} \frac{M_{p a}^{m} M_{q a}^{m}}{\omega - \Omega_m - \epsilon_a + \imath \eta}
\end{equation}
where we have introduced the elements of the transition densities
\begin{equation}
        M_{pq}^{m} = \sum_{ia} \braket{pi}{qa} \qty( X_{ia}^{m} + Y_{ia}^{m} )
\end{equation}
The braket notation is employed to represent the bare two-electron Coulomb integrals
\begin{equation}
	\braket{pq}{rs} = \iint \frac{\varphi_p(\bx{1}) \varphi_q(\bx{2}) \varphi_r(\bx{1}) \varphi_s(\bx{2})}{\left|\br{1} - \br{2}\right|} d\bx{1} d\bx{2}
\end{equation}
The excitation energies $\Omega_m$ and amplitudes $X_{ia}^{m}$, $Y_{ia}^{m}$ are obtained as eigenvalues and eigenvectors of a Casida-like eigenproblem
\begin{equation}
	\mqty(\boldsymbol{A} & \boldsymbol{B}  \\ -\boldsymbol{B} & -\boldsymbol{A}) \, \mqty(\boldsymbol{X}_{m} \\ \boldsymbol{Y}_{m})
	= \Omega_m \, \mqty(\boldsymbol{X}_{m} \\ \boldsymbol{Y}_{m})
\end{equation}
where
\begin{equation}
\begin{aligned}
	A_{ia,jb} &= (\epsilon_a - \epsilon_i) \delta_{ij} \delta_{ab} + \braket{ib}{aj} \\
	B_{ia,jb} &= \braket{ij}{ab}
\end{aligned}
\end{equation}

We now shift our focus to the computation of the Dyson orbitals $\{\varphi_p\}$ and quasiparticle energies $\{\epsilon_p\}$. 
The Dyson equation~\eqref{eq:G_Dyson} implies that these quantities should satisfy the following dynamical non-Hermitian equation
\begin{equation}
	\qty[ F + \Sigma_c\qty(\omega=\epsilon_p) ] \varphi_p = \epsilon_p \, \varphi_p
	\label{eq:qp_eqt}
\end{equation}
where $F$ is the Fock operator.
However, the frequency dependence of the correlation self-energy $\Sigma_c$ introduces complexity and non-linearity to this quasiparticle equation. 
As a result, various common approximations are employed.
The widely-used $G_0 W_0$ scheme involves a single-shot iteration of Eq.~\eqref{eq:qp_eqt}, considering only the diagonal part of 
the self-energy. \cite{Strinati_1980,Hybertsen_1985a,Godby_1988,Linden_1988,Northrup_1991,Blase_1994,Rohlfing_1995} For instance, starting with a set of one-electron HF orbitals $\{\varphi_p^{\text{HF}}\}$, where 
$F \varphi_p^{\text{HF}} = \epsilon_p^{\text{HF}} \varphi_p^{\text{HF}}$, the following equations are obtained and solved
\begin{equation}
	\epsilon_p^{\text{HF}} + \left[\Sigma_c\qty(\omega)\right]_{pp} = \omega
	\label{eq:qp_eqt_diag}
\end{equation}
Linearizing this equation is a common practice, achieved by performing a first-order Taylor expansion of the self-energy around $\omega=\epsilon_p^{\text{HF}}$.
An iterative approach, known as ev$GW$, goes a step further by updating the eigenvectors $\boldsymbol{X}_m$, $\boldsymbol{Y}_m$, and eigenvalues $\Omega_m$, and consequently updating the self-energy, until 
convergence over the quasiparticle energies $\{\epsilon_p\}$ is achieved.\cite{Hybertsen_1986,Shishkin_2007a,Blase_2011,Faber_2011,Rangel_2016}
The qs$GW$ method introduces another level of self-consistency, where both orbitals and energies are iteratively updated until convergence. \cite{Gui_2018,Faleev_2004,vanSchilfgaarde_2006,Kotani_2007,Ke_2011,Kaplan_2016,Forster_2021,Marie_2023}
However, to avoid dealing with the non-Hermitian and dynamical nature of the correlation self-energy, a static symmetric approximation is considered instead, which reads
\begin{equation}
	\mel{\varphi_p}{F}{\varphi_q} + 
	\frac{[\Sigma_c\qty(\epsilon_p)]_{pq} + [\Sigma_c\qty(\epsilon_q)]_{qp}}{2} 
	= \epsilon_p \, \delta_{pq}
	\label{eq:qp_qs}
\end{equation}
Recently, a qs$GW$ scheme based on a static Hermitian self-energy obtained from a similarity renormalization group approach has been proposed as an alternative to Eq.~\eqref{eq:qp_qs}.\cite{Marie_2023}

Based on these calculations, the principal IP and EA of a given system are obtained as
\begin{align}
	\IP & = - \epsilon_{\HOMO}
	&
	\EA & = - \epsilon_{\LUMO}
\end{align}
where HOMO and LUMO are the highest-occupied and lowest-unoccupied molecular orbitals, respectively.
These identities are valid at the HF and $GW$ levels.

%%%%%%%%%%%%%%%%%%%%%%%%%%%%%%%
\section{Computational details}
%%%%%%%%%%%%%%%%%%%%%%%%%%%%%%%
The reference data specifically produced for the present study have been obtained at the full CI (FCI) level. All these calculations have been performed with \textsc{quantum package} \cite{Garniron_2019} using the \textit{``Configuration Interaction using a Perturbative Selection made Iteratively''} (CIPSI) algorithm \cite{Huron_1973,Giner_2013,Giner_2015,Garniron_2017,Garniron_2018} and within the frozen-core approximation.

The $GW$ calculations have been carried out with \textsc{quack}, an open-source software for emerging quantum electronic structure methods, for which the source code is available at \url{https://github.com/pfloos/QuAcK}. Their algorithm and implementation are described in Ref.~\onlinecite{Marie_2023b}. In the $G_0W_0$ and ev$GW$ calculations, we set $\eta = 0$ and we solve the frequency-dependent quasiparticle equation without relying on its linearization to get the quasiparticle energies. The qs$GW$ calculations are performed with the regularized scheme based on the similarity renormalization group approach, as mentioned above and described in Ref.~\onlinecite{Marie_2023}. A flow parameter of $s = 1000$ is employed. All (occupied and virtual) orbitals are corrected.

The systems considered here have a closed-shell electronic structure and, unless otherwise stated, we have opted to maintain spatial and spin symmetry. Therefore, we rely on the restricted formalism for the HF and $GW$ calculations. The restricted HF (RHF) calculations are systemically initiated with a core Hamiltonian guess and an internal stability analysis of the RHF solution towards other RHF solutions is systematically performed. \cite{Seeger_1977,Fukutome_1981,Stuber_2003} All $GW$ calculations employed these RHF quantities as a starting point. A systematic treatment of these systems with more exotic HF formalisms, including the unrestricted and/or generalized approaches, is deferred to future work. \cite{Shirley_1993,Yamanaka_2001,Mansouri_2021,Pokhilko_2022b,Pokhilko_2023}
For each system and method, the raw data are collected in the \SupMat.

%%%%%%%%%%%%%%%%%
\section{Results}
%%%%%%%%%%%%%%%%%

%=================================
\subsection{\ce{Be + H2} reaction}
%=================================

%%% FIG. 1 %%%
\begin{figure*}
	\includegraphics[width=0.33\linewidth]{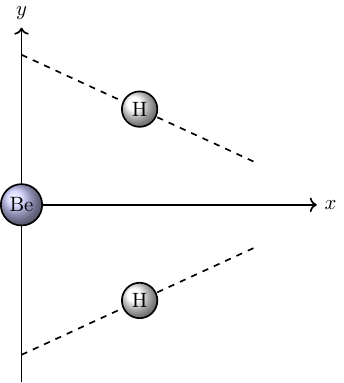}
	\includegraphics[width=0.66\linewidth]{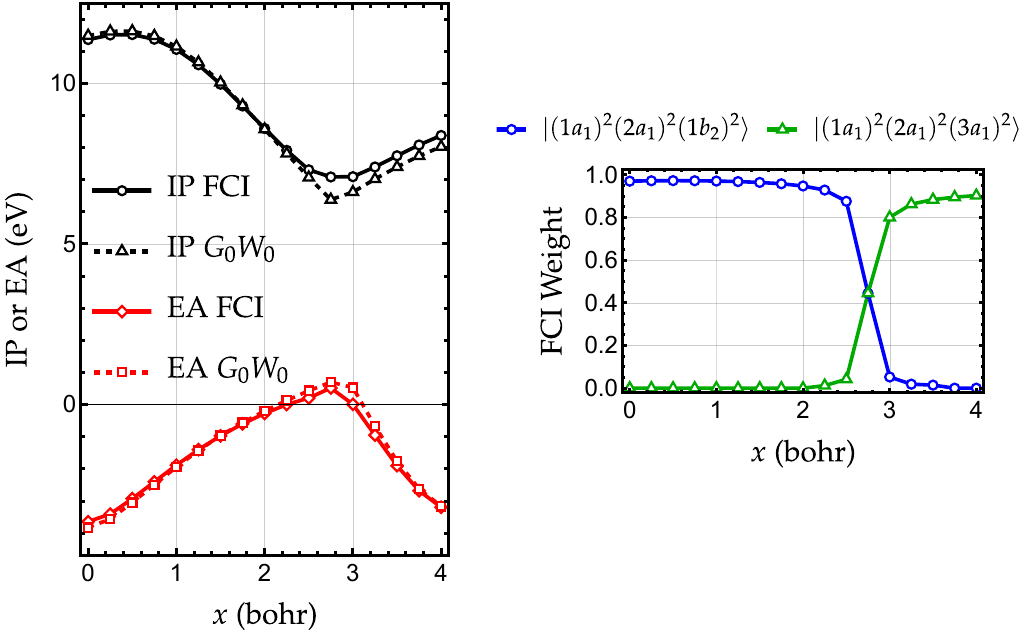}
	\caption{Left: Sketch of the insertion reaction of \ce{Be} into \ce{H2}. The $x$ coordinates varies from 0 to 4 bohr, and $y = 2.54 - 0.46x$. Center: Variations of the principal IP and EA (in \si{\eV}) during the reaction as functions of $x$. Right: Evolution of the FCI weights associated with the two dominant electronic configurations, $\ket*{(1a_1)^2 (2a_1)^2 (1b_2)^2}$ and $\ket*{(1a_1)^2 (2a_1)^2 (3a_1)^2}$, as functions of $x$.}
	\label{fig:sketch_BeH2}
\end{figure*}

%%% FIG. 2 %%%
\begin{figure}
	\includegraphics[width=\linewidth]{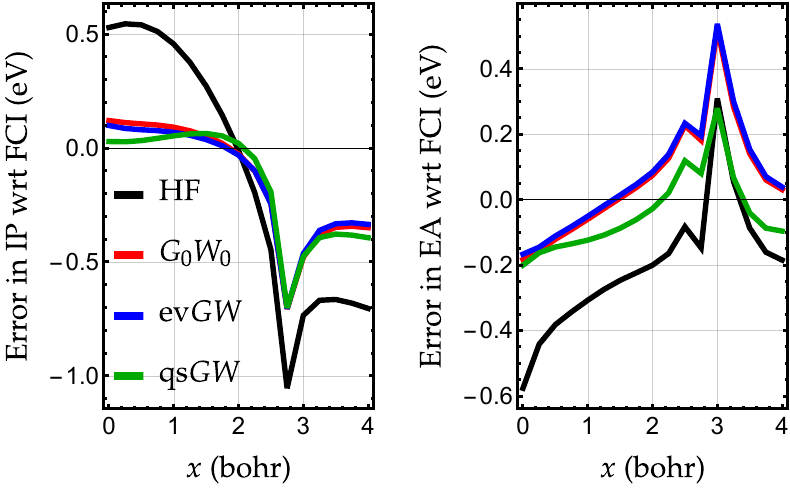}
	\caption{Error in IP and EA with respect to FCI (in \si{\eV}) during the insertion reaction of \ce{Be} into \ce{H2} as a function of $x$ computed at the HF, $G_0W_0$, ev$GW$, and qs$GW$ levels.}
	\label{fig:BeH2}
\end{figure}

Using a simple Be(3s2p)/H(2s) basis set (see \SupMat) and correlating all electrons, we initially examine the insertion of a beryllium atom into \ce{H2} to form \ce{BeH2} following a C$_\text{2v}$ pathway, \cite{Purvis_1983} or at least the variant proposed by Evangelista and coworkers. \cite{Evangelista_2011a,Evangelista_2011b,Evangelista_2012} As depicted in the left panel of Fig.~\ref{fig:sketch_BeH2}, the \ce{Be} atom is placed at the center of the coordinate system, and two hydrogen atoms are located at $(x,\pm y,0)$ with $y = 2.54 - 0.46x$ and $x$ ranging from $0$ to $4$ bohr. At small $x > 0$, the FCI wave function of \ce{BeH2} is dominated by the electronic configuration $\ket{(1a_1)^2 (2a_1)^2 (1b_2)^2}$, while for larger $x$, the configuration $\ket{(1a_1)^2 (2a_1)^2 (3a_1)^2}$ prevails. In the region $2.5 < x < 3$, the wave function switches rapidly from $\ket{(1a_1)^2 (2a_1)^2 (1b_2)^2}$ to $\ket{(1a_1)^2 (2a_1)^2 (3a_1)^2}$, as illustrated in the right panel of Fig.~\ref{fig:sketch_BeH2}. Particularly, at $x = 2.75$, the wave function contains an equal amount of the two configurations. This region of strong multireference effects is anticipated to produce the largest deviations between the reference FCI values and the $GW$-based methods.

To determine the exact IP and EA of this system, we performed FCI calculations on the cation, neutral, and anionic species at each geometry from $x=0$ to $x=4$ (central panel of Fig.~\ref{fig:sketch_BeH2}). Starting from $x=0$, the IP decreases, reaching a minimum at $x=2.75$, while the EA increases to its maximum at the same point. Additionally, we computed the errors in IP and EA (with respect to FCI) across the entire range of $x$ values using HF, $G_0W_0$, ev$GW$, and qs$GW$ (see Fig.~\ref{fig:BeH2}). Except for $x = 2.75$, the RHF solution remains internally stable (i.e., there is no RHF-to-RHF instability). At $x = 2.75$, one can find a spatially broken-symmetry RHF solution with energy marginally lower than the symmetry-pure RHF solution. The numerical results obtained with these two solutions are extremely close, and we have thus chosen to consider the symmetry-pure solution in the following.

At each level of theory, the behavior of the charged excitation energies closely mirrors the FCI results, except around $x=2.75$, where a significant deviation occurs. At the FCI level, the transition between the two regions is smooth, while at HF level and hence at the $GW$ level as well, the transition is better described by two solutions crossing abruptly. This behavior is ubiquitous in excited-state HF calculations.\cite{Burton_2015,Burton_2018,Marie_2021} Overall, Fig.~\ref{fig:BeH2} illustrates that $GW$ notably improves upon HF, with $G_0W_0$ and ev$GW$ exhibiting close agreement. Besides, qs$GW$ only improves in the small-$x$ region and does not provide more accurate properties in the problematic region around $x=2.75$, where all $GW$ methods yield essentially the same values. In the worst-case scenario, $GW$ deviates by \SI{0.7}{\eV} for the IP and \SI{0.5}{\eV} for the EA. Furthermore, the isolated \ce{Be} also features a strong competition between the $\ket{(1s)^2(2s)^2}$ and $\ket{(1s)^2(2p)^2}$ configurations. \cite{Mok_1996} Therefore, the large-$x$ limit of \ce{Be + H2} will also have multireference characteristics, which are evident in the larger errors in the predicted IP for $x > 2.75$. These results highlight that the quality of the HF reference wave function is crucial and that self-consistency does not lead to any significant improvement. Nonetheless, we can conclude that the $GW$ approximation provides a quantitative description of the \ce{Be + H2} reaction, except in the strongly multireference region where the agreement is only qualitative.

%==================================
\subsection{Multireference systems}
%==================================

%%% TABLE I %%%
\begin{table}
	\caption{Ground-state geometry (in \si{\angstrom} and degree) of the multireference systems considered herein, as well as the weight of the reference configuration in the FCI wave function for the cationic, neutral, and anionic singlet ground states.}
\label{tab:geom}
\begin{ruledtabular}
\begin{tabular}{llccc}
				&      							&  \mc{3}{c}{Reference Weight}	\\
															\cline{3-5}
	System		&	Geometry					&	Cation	&	Neutral	&	Anion	\\
	\hline
	\ce{B2}		&   $R_{\ce{BB}} = 1.59$		&	0.73	&	0.36	&	0.71	\\
	\ce{LiF}	&	$R_{\ce{LiF}} = 1.5639$		&	0.96	&	0.93	&	0.94	\\
	\ce{BeO}	&	$R_{\ce{BeO}} = 1.3308$		&	0.93	&	0.90	&	0.94	\\
	\ce{BN}		&	$R_{\ce{BN}} = 1.281$		&	0.69	&	0.69	&	0.80	\\
	\ce{C2}		&	$R_{\ce{CC}} = 1.2425$		&	0.69	&	0.69	&	0.82	\\
	\ce{O3}		&  	$R_{\ce{OO}} = 1.278$ 		&	0.74	&	0.76	&	0.76	\\
				&  	$\angle_{\ce{OOO}} = 116.8$	&			&			&			\\
\end{tabular}
\end{ruledtabular}
\end{table}

%%% TABLE II %%%
\begin{table*}
	\caption{Principal IP, principal EA, and fundamental gap (in \si{\eV}) for a selection of multireference systems computed at the $G_0W_0$, ev$GW$, qs$GW$, and FCI levels of theory with the def2-TZVPP basis. \#1 and \#2 correspond to distinct RHF solutions whose properties are collected in Table \ref{tab:HF}. The error with respect to the reference FCI value is reported in parentheses.}
\label{tab:res_MR}
	\begin{ruledtabular}
		\begin{tabular}{llrrrrrrd}
	 	&		&	\mc{2}{c}{$G_0W_0$} 		&	\mc{2}{c}{ev$GW$} 			&	\mc{2}{c}{qs$GW$}			\\	
					\cline{3-4}						\cline{5-6}						\cline{7-8}		
Mol. 	&		&	\tabc{\#1} 		&	\tabc{\#2}		&	\tabc{\#1} 	&	\tabc{\#2}	&	\tabc{\#1} 	&	\tabc{\#2}	&	\tabc{FCI} \\					
		\hline
\ce{B2}	&	IP	&	9.06($+0.09$)	&	8.87($-0.10$)	&	9.10($+0.13$)	&	8.87($-0.10$)	&	8.84($-0.13$)	&	8.71($-0.26$)	&	8.97	\\
	 	&	EA	&	2.05($+0.19$)	&	2.13($+0.27$)	&	2.12($+0.26$)	&	2.20($+0.34$)	&	1.90($+0.04$)	&	2.15($+0.29$)	&	1.86	\\
	 	&	Gap	&	7.01($-0.11$)	&	6.74($-0.37$)	&	6.98($-0.14$)	&	6.69($-0.42$)	&	6.94($-0.18$)	&	6.56($-0.55$)	&	7.11	\\
\\
\ce{LiF}&	IP	&	11.31($-0.01$)	&					&	11.09($-0.23$)	&					&	11.38($+0.06$)	&					&	11.32	\\
		&	EA	&	0.01($-0.01$)	&					&	0.02($+0.00$)	&					&	0.00($-0.02$)	&					&	0.02	\\
		&	Gap	&	11.29($-0.01$)	&					&	11.07($-0.23$)	&					&	11.38($+0.08$)	&					&	11.30	\\
\\
\ce{BeO}&	IP	&	9.76($-0.21$)	&					&	9.63($-0.34$)	&					&	10.10($+0.13$)	&					&	9.97	\\
		&	EA	&	2.09($+0.12$)	&					&	2.12($+0.15$)	&					&	1.94($-0.03$)	&					&	1.97	\\
		&	Gap	&	7.67($-0.32$)	&					&	7.51($-0.48$)	&					&	8.17($+0.17$)	&					&	8.00	\\
\\
\ce{BN}	&	IP	&	11.69($-0.24$)	&	11.90($-0.03$)	&	11.68($-0.24$)	&	11.93($-0.03$)	&	11.83($-0.10$)	&	11.83($-0.10$)	&	11.93	\\
		&	EA	&	3.83($+0.84$)	&	3.70($+0.71$)	&	3.89($+0.90$)	&	3.76($+0.77$)	&	3.34($+0.35$)	&	3.34($+0.35$)	&	2.99	\\
		&	Gap	&	7.86($-1.08$)	&	8.20($-0.73$)	&	7.79($-1.15$)	&	8.17($-0.79$)	&	8.49($-0.45$)	&	8.49($-0.45$)	&	8.94	\\
\\
\ce{C2}	&	IP	&	12.92($+0.48$)	&	12.42($-0.03$)	&	12.95($+0.50$)	&	12.42($-0.03$)	&	12.54($+0.09$)	&	12.16($-0.29$)	&	12.45	\\
		&	EA	&	4.08($+1.08$)	&	4.40($+1.40$)	&	4.16($+1.16$)	&	4.48($+1.48$)	&	3.73($+0.73$)	&	4.40($+1.40$)	&	3.00	\\
	 	&	Gap	&	8.85($-0.60$)	&	8.02($-1.43$)	&	8.79($-0.66$)	&	7.93($-1.52$)	&	8.81($-0.64$)	&	7.76($-1.69$)	&	9.45	\\
\\
\ce{O3}	&	IP	&	13.50($+0.92$)	&					&	13.40($+0.82$)	&					&	13.12($+0.54$)	&					&	12.58	\\
		&	EA	&	1.96($+0.68$)	&					&	2.01($+0.73$)	&					&	1.85($+0.57$)	&					&	1.28	\\
		&	Gap	&	11.54($+0.25$)	&					&	11.39($+0.10$)	&					&	11.28($-0.01$)	&					&	11.29	\\
		\end{tabular}
	\end{ruledtabular}
\end{table*}

%%% TABLE III %%%
\begin{table}
	\caption{Properties of the two RHF solutions of \ce{B2}, \ce{BN}, and \ce{C2} computed with the def2-TZVPP basis. The negative eigenvalues (in \si{\hartree}) of the internal stability analysis are reported alongside the RHF energy (in \si{\hartree}), $E_{\HF}$, and the HOMO and LUMO orbital energies (in \si{\eV}), $\e_{\HOMO}^{\HF}$ and $\e_{\LUMO}^{\HF}$.}
	\label{tab:HF}
	\begin{ruledtabular}
	\begin{tabular}{lccccc}
	Mol.		&	Sol.	&	$E_{\HF}$		&	Int. Stab.	&	$\e_{\HOMO}^{\HF}$	&	$\e_{\LUMO}^{\HF}$	\\
	\hline
	\ce{B2}		&	\#1 	&	$-49.042 173$	&	$-0.043$	&	$-8.54$		&	$-1.18$	\\
				&	\#2 	&	$-49.059 358$	&				&	$-8.71$		&	$-1.05$	\\
	\\
	\ce{BN}		&	\#1 	&	$-78.908 465$	&	$-0.019$	&	$-11.53$	&	$-2.93$	\\	
				&	\#2 	&	$-78.911 128$	&				&	$-11.16$	&	$-2.69$	\\
	\\
	\ce{C2}		&	\#1 	&	$-75.403 580$	&	$-0.067$	&	$-12.46$	&	$-3.12$	\\	
				&	\#2 	&	$-75.439 770$	&				&	$-12.79$	&	$-2.78$	\\
	\end{tabular}
	\end{ruledtabular}
\end{table}

In the second stage of this study, we explore a set of molecules exhibiting varying degrees of multireference character at their respective experimental equilibrium geometry.  \cite{HerzbergBook} The geometric parameters for these molecules are compiled in Table \ref{tab:geom}. The respective weights of the RHF reference determinant in the cationic, neutral, and anionic singlet ground-state wave functions are reported in the same table. Here, we employ the more realistic triple-$\zeta$ basis set, def2-TZVPP. \cite{Weigend_2005}

The boron dimer displays a pronounced multireference character in its lowest singlet state (HF determinant weight of only $0.36$) although the true ground state possesses triplet spin symmetry ($^3{}\Sigma_\text{g}^{-}$), and the corresponding HF wave function has the configuration $\ket*{\cdots \sigma_\text{u}^2 \sigma_\text{g}^2 \sigma_\text{u}^2 \pi_\text{u}^2}$.\cite{Graham_1976,Dupuis_1978,Deutsch_1990,Bruna_1990} The carbon dimer serves as a prototypical multireference system (weight of $0.69$ on the reference determinant) extensively studied in the literature using state-of-the-art electronic structure methods.\cite{Bauschlicher_1987,Abrams_2004,Sherrill_2005,Li_2006,Booth_2011} We also consider other members of the 12-electron series (\ce{LiF}, \ce{BeO}, and \ce{BN} \footnote{Although we consider the lowest-energy singlet state, it is worth noting that \ce{BN} has a triplet ground state. \cite{Lorenz_1996}}) and ozone, which are all part of the $GW100$ dataset. \cite{vanSetten_2015} Moving from \ce{LiF} to \ce{C2}, the multireference character magnifies. Notably, all these systems exhibit a positive electron affinity, implying the stability of their corresponding anion. 

Our results, summarized in Table \ref{tab:res_MR}, include the IP, EA, and fundamental gap computed at the $G_0W_0$, ev$GW$, qs$GW$, and FCI levels of theory. Errors with respect to the reference FCI values are indicated in parentheses. Starting with a core guess, for the more pronounced multireference systems (\ce{B2}, \ce{BN}, and \ce{C2}), an internally unstable RHF solution (labeled as \#1) is obtained. By following the eigenvector associated with the negative eigenvalue, a lower-lying RHF solution labeled as \#2 is reached (see Table \ref{tab:HF}). For \ce{B2}, both RHF solutions have broken spatial symmetry. In \ce{C2}, \#1 has a configuration $\ket*{\cdots \sigma_\text{g}^2 \sigma_\text{u}^2 \pi_\text{u}^2 \pi_\text{u}^2}$ and is of $^1{}\Sigma_\text{g}^{+}$ symmetry, while \#2 has broken spatial symmetry. A similar situation arises in \ce{BN}, where \#1 has $^1{}\Sigma^+$ symmetry and a configuration $\ket*{\cdots \sigma^2 \sigma^2 \pi^2 \pi^2}$, while the wave function of \#2 is spatially broken. For \ce{LiF}, \ce{BeO}, and \ce{O3}, \#1 is found to be internally stable. The IP and EA obtained with $GW$ based on \#1 are very acceptable, but in some cases, improvement can be obtained by considering the lower-energy RHF solution \#2, as explained further below.

For the IP of \ce{B2}, transitioning from \#1 to \#2 results in a negative shift of approximately \SI{0.2}{\eV}. At the $G_0W_0$ and ev$GW$ levels, the exact result falls almost exactly between the quasiparticle energies obtained with the two reference RHF solutions. Consequently, we observe a very small improvement with errors around \SI{0.1}{\eV}, unexpectedly well below the mean absolute error of $GW$ calculated on the $GW100$ benchmark test set (\SI{0.31}{\eV} at the $G_0W_0$@HF level). Due to self-consistency overcorrecting the IPs, the error in qs$GW$ reaches \SI{0.26}{\eV} when considering \#2 as the starting point. The errors on the EAs are larger, as expected. Nonetheless, the overall trend is very similar.

It is noteworthy that although qs$GW$ is generally considered independent of the starting point due to self-consistency on quasiparticle energies and corresponding orbitals, initiating the qs$GW$ self-consistent process with \#1 or \#2 may lead to different sets of results. As qs$GW$ considers a Fock-like operator including Hartree, exchange, and correlation (similar to Kohn-Sham calculations), it is not surprising to locate different solutions at the qs$GW$ level. In the case of the carbon dimer, using \#2 as a starting point significantly improves the IPs for $G_0W_0$ and ev$GW$, while it deteriorates the qs$GW$ results. Similar observations apply to \ce{BN}. In particular, considering the lowest-energy RHF solution significantly improves both the IP and EA computed at the $G_0W_0$ and ev$GW$ levels. It is interesting to note that, in the case of \ce{BN}, starting the qs$GW$ calculations with \#1 or \#2 leads to the same quasiparticle energies, and again the accuracy reached is quite reasonable.

For the more weakly correlated systems, \ce{LiF} and \ce{BeO}, the IPs and EAs are well reproduced by $GW$, especially at the $G_0W_0$ and qs$GW$ levels (ev$GW$ errors are slightly larger). For \ce{O3}, which exhibits a more significant multireference character than the two previous systems, errors are larger (up to almost \SI{1}{\eV} for $G_0W_0$). Nonetheless, the qs$GW$ formalism can significantly reduce these errors. In conclusion, the best compromise appears to be qs$GW$@RHF using the symmetry-pure solution (\#1), providing accurate IPs and EAs for the weakly and more strongly correlated systems that are considered here.

For the sake of completeness, we report, in the \SupMat, $G_0W_0$ results computed with Kohn-Sham starting points (BLYP, \cite{Becke_1988,Lee_1988} B3LYP, \cite{Becke_1988,Lee_1988,Becke_1993a} and CAM-B3LYP\cite{Yanai_2004}) for the same set of molecules. The stability analysis reveals that all the considered density-functional approximations lead to a unique stable restricted solution and that only $G_0W_0$@CAM-B3LYP produces competitive results when compared to $G_0W_0$@HF.

%=============================================
\subsection{Triangular-shaped \ce{H6} cluster}
%=============================================

%%% TABLE IV %%%
\begin{table*}
	\caption{Nature and properties of the different HF solutions located for the triangular-shaped \ce{H6} cluster using the STO-6G basis. The HF energy (in \si{\hartree}), $E_{\HF}$, and the expectation value of $\hS^2$ are indicated for each solution. The IP and EA (in \si{\eV}) computed at the HF, $G_0W_0$, qs$GW$, and FCI levels are also reported and the error with respect to the reference FCI value is indicated in parentheses.}
	\label{tab:H6}
	\begin{ruledtabular}
	\begin{tabular}{llrrrrrrrrrr}
				&	&						&								&	\mc{2}{c}{HF}				&	\mc{2}{c}{$G_0W_0$}			&	\mc{2}{c}{qs$GW$}			&	\mc{2}{c}{FCI}				\\
																		\cline{5-6}						\cline{7-8}						\cline{9-10}		\cline{11-12}
	Nat.	&	Sol.	&	\tabc{$E_{\HF}$}	&	\tabc{$\expval*{\hS^2}$}	&	\tabc{IP}	&	\tabc{EA}	&	\tabc{IP}	&	\tabc{EA}	&	\tabc{IP}	&	\tabc{EA}	&	\tabc{IP}	&	\tabc{EA}	\\
	\hline
	RHF	&			&	$-2.449 047$	&	$0.000$	&	$6.36(-2.91)$	&	$-3.22(+2.01)$	&	$6.55(-2.72)$	&	$-2.94(+2.29)$	&	$6.48(-2.79)$	&	$-2.86(+2.37)$	&	$9.27$	&	$-5.23$	\\
	UHF	&	\#1		&	$-2.798 321$	&	$2.821$	&	$10.38(+1.11)$	&	$-5.51(-0.28)$	&	$10.09(+0.82)$	&	$-5.28(-0.05)$	&	$10.01(+0.74)$	&	$-5.20(+0.03)$	&	$9.27$	&	$-5.23$	\\
	UHF	&	\#2		&	$-2.819 463$	&	$2.742$	&	$10.42(+1.15)$	&	$-6.92(-1.69)$	&	$10.06(+0.79)$	&	$-6.56(-1.33)$	&	$9.95(+0.68)$	&	$-6.43(-1.20)$	&	$9.27$	&	$-5.23$	\\
	UHF	&	\#3		&	$-2.824 460$	&	$2.712$	&	$11.16(+1.89)$	&	$-6.26(-1.03)$	&	$10.75(+1.48)$	&	$-5.94(-0.71)$	&	$10.61(+1.34)$	&	$-5.84(-0.61)$	&	$9.27$	&	$-5.23$	\\
	\end{tabular}
	\end{ruledtabular}
\end{table*}

%%% FIG. 3 %%%
\begin{figure*}
	\includegraphics[width=0.8\linewidth]{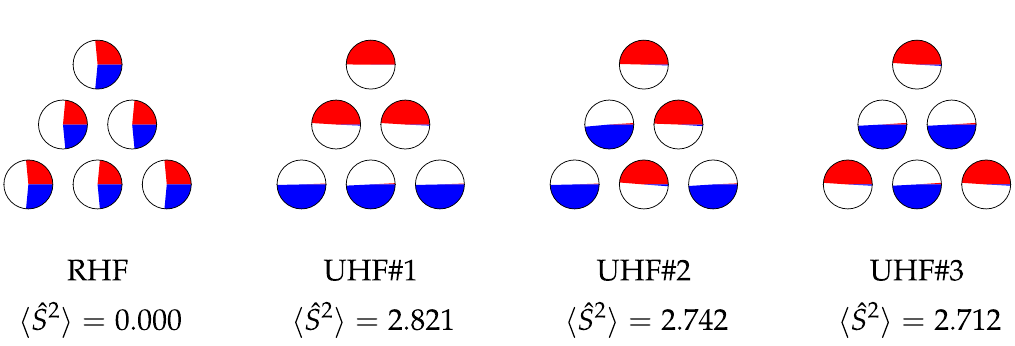}
	\caption{Mulliken population analysis and the expectation values of $\hS^2$ for the four HF solutions of the triangular-shaped \ce{H6} cluster (see Table \ref{tab:H6}). A full half-circle corresponds to an entire spin-up or spin-down electron.}
	\label{fig:H6}
\end{figure*}

The geometry of the \ce{H6} cluster, with a \SI{2}{\angstrom} separation between each hydrogen atom, is reported in the \SupMat. To simplify the present analysis, we employ the minimal STO-6G basis set. At the FCI/STO-6G level, the ground-state energies for the cation, neutral, and anionic species are \SI{-2.516 380}{\hartree}, \SI{-2.857 023}{\hartree}, and \SI{-2.664 959}{\hartree}, respectively, resulting in an IP and EA of \SI{9.27}{\eV} and \SI{-5.23}{\eV}, respectively. The RHF estimates of the IP and EA deviate significantly from these FCI values, with an offset of approximately \SI{3}{\eV} for the IP and \SI{2}{\eV} for the EA. Moreover, performing a $GW$ calculation on top of the RHF results does not yield any improvement, and self-consistency has minimal impact, slightly worsening the results.

While the RHF solution is internally stable, it is unstable towards UHF (RHF-to-UHF instabilities). The stability analysis reveals five negative eigenvalues: one non-degenerate at \SI{-0.647}{\hartree} and two sets of doubly degenerate eigenvalues at \SI{-0.362}{\hartree} and \SI{-0.079}{\hartree}. Following the lowest eigenvalues leads to the lowest-energy (stable) solution, UHF\#3, while each pair of degenerate eigenvalues leads to distinct stable solutions, UHF\#1 and UHF\#2 (see Table \ref{tab:H6}). The significant spin contamination of these UHF solutions, as indicated by the values of $\expval*{\hS^2}$, reveals the spin frustration inherent in this system.

Figure \ref{fig:H6} illustrates the Mulliken population analysis of the four stable HF solutions we have identified. 
The spin-$\sigma$ electronic population on nucleus $A$ is given by \cite{SzaboBook} 
\begin{equation}
	q_A^\sigma = - \sum_{\mu \in A} (\boldsymbol{P}^\sigma \cdot \boldsymbol{S})_{\mu\mu}
\end{equation}
where $\boldsymbol{P}^\sigma$ is the spin-$\sigma$ density matrix and $\boldsymbol{S}$ is the overlap matrix, both expressed in the atomic orbital basis.
A full half-circle represents an entire spin-up or spin-down electron located on this atom. In the RHF wave function, the spin-up and spin-down populations are identical, and the distribution of electrons on each site is nearly equal. However, in the UHF wave functions, electrons localize on specific centers, creating different patterns. 

Transitioning from RHF to UHF results in a drastic improvement in the IP and EA estimates, highlighting the practical impact of breaking spatial and spin symmetries in the presence of spin frustration. However, determining which solution to favor in this case remains unclear: the lowest-energy solution with the largest spin contamination, UHF\#3, is clearly inferior to the two others, while UHF\#2 is slightly better for IPs but yields poor estimates of the EAs compared to UHF\#1.
These results highlight the practical challenges faced by $GW$ calculations for molecules with severe multireference effects.

%=============================================
\subsection{Dissociation of \ce{HF}}
%=============================================

%%% FIG. 4 %%%
\begin{figure}
	\includegraphics[width=\linewidth]{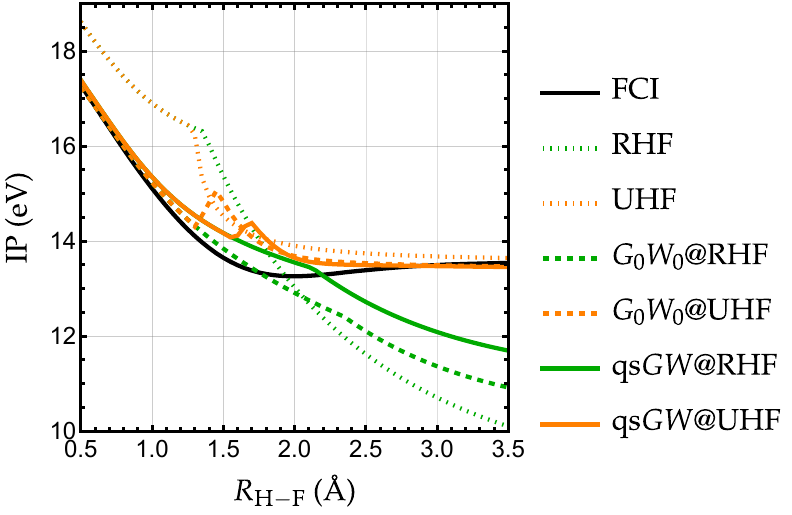}
	\caption{Variation of the IP (in \si{\eV}) of the \ce{HF} molecule during its dissociation computed at various levels of theory. Restricted- and unrestricted-based calculations are represented in green and orange, respectively.}
	\label{fig:HF}
\end{figure}

Our final example deals with the dissociation of the \ce{HF} molecule. Using Dunning's cc-pVDZ basis set, we compute the IP as a function of the internuclear distance $R_{\ce{H-F}}$ ranging from \SIrange{0.5}{3.5}{\angstrom} at the FCI, HF, $G_0W_0$, and qs$GW$ levels. Our results are reported in Fig.~\ref{fig:HF}. Here again, we employ both (stable) RHF and UHF starting points for the $G_0W_0$ and qs$GW$ calculations. At equilibrium, the dominant closed-shell configuration is $\ket*{\cdots(3\sigma)^2(1\pi)^4}$ while, at stretched geometries, two additional configurations, $\ket*{\cdots(1\pi^4)(4\sigma)^2}$ and $\ket*{\cdots(1\pi^4)(3\sigma)(4\sigma)}$, must be considered.

The FCI curve is rather simple: the IP decreases from a value of approximately \SI{17}{\eV} at $R_{\ce{H-F}} = \SI{0.5}{\angstrom}$ to reach a minimum of \SI{13.3}{\eV} around $R_{\ce{H-F}} = \SI{2}{\angstrom}$ before slightly increasing toward a limiting value of approximately \SI{13.6}{\eV} for large bond lengths. For small $R_{\ce{H-F}}$, the RHF IP is a rather poor approximation of its FCI counterpart but follows the correct trend. In this regime, the $G_0W_0$ perturbative correction is effective up to $\SI{1.5}{\angstrom}$. However, beyond this point, it deviates significantly. As observed previously, qs$GW$ does not match the accuracy of $G_0W_0$ but follows a similar trend.

A UHF solution emerges around $\SI{1.4}{\angstrom}$ and the UHF IP provides an excellent estimate of the FCI value for large bond lengths, with a systematic improvement at both the $G_0W_0$ and qs$GW$ levels. However, the transition between the RHF and UHF starting points results in non-smooth curves at the $GW$ level, featuring bumps around the Coulson-Fischer point. Although this bump is mitigated at the qs$GW$ level due to the self-consistency over the orbitals, it remains present. In this region, the self-consistent qs$GW$ calculations are quite difficult to converge hinting at the presence of multiple solutions. \cite{Loos_2018b,Veril_2018,Pokhilko_2021a,Pokhilko_2021b,Berger_2021,DiSabatino_2021,Pokhilko_2022a,Monino_2022} Apart from this, the qs$GW$@UHF curve accurately follows the FCI dissociation curve, providing a rather satisfactory description of this single-bond dissociation process.

%%%%%%%%%%%%%%%%%%%%%%%%%%%%%%%%%%%%%
\section{Conclusion and Perspectives}
%%%%%%%%%%%%%%%%%%%%%%%%%%%%%%%%%%%%%
The present study highlights the diverse behavior of $GW$ in the presence of strong correlation. 
For the \ce{Be + H2} insertion reaction, the $GW$ approximation provides a quantitative description, except in regions of strong multireference character, where the agreement is merely qualitative. For molecules with varying amounts of multireference character, the optimal compromise emerges with qs$GW$ employing a symmetry-pure HF reference whenever available. This approach yields IP and EA estimates for both weakly and strongly correlated systems with notable accuracy. In contrast, the spin-frustrated \ce{H6} cluster in a triangular arrangement and the dissociation of \ce{HF} reveal that breaking spin symmetry is wise and useful in certain contexts. For the \ce{H6} cluster, the RHF-based estimates exhibit significant deviations, whereas IPs and EAs obtained through qs$GW$ with a UHF reference align much more closely with the FCI reference values. However, because various UHF solutions do exist, it is unclear which solution to favor in this case.
Furthermore, the dissociation of the \ce{HF} molecule demonstrates that self-consistency in addition to symmetry breaking can be useful for single-bond breaking processes, although the dissociation curve exhibits an unphysical ``bump''  near the Coulson-Fischer point. Scenarios involving multiple-bond breaking remain to be studied in this context.

Notably, the discrepancy in accuracy for different variants of $GW$ and initial states is most pronounced for molecules that undergo spin-symmetry breaking at the Hartree-Fock level, including the spin-frustrated \ce{H6} cluster and the dissociation of \ce{HF}, and is relatively less severe in the two-configuration scenario of \ce{BeH2}. To overcome the challenges for spin-frustrated systems, we intend to explore the generalized version of $GW$ which allows the $\hS_z$ symmetry to be broken, thus enabling the use of non-collinear reference wave functions for subsequent $GW$ post-treatment.
Strikingly, the accuracy remains acceptable in the case of \ce{B2}, despite its strong multireference character. This somewhat surprising result may be attributed to error compensation arising from the interplay between the level of self-consistency, the precise nature of the underlying mean-field solution, and the degree of multireference character.

We set out to investigate the accuracy of the $GW$ approximation for multireference systems. Our findings indicate that, indeed, $GW$ can describe such systems to a certain extent. However, it is clear that the errors are notably larger than those encountered in single-reference systems. While the precise relationship between the magnitude of this error and the number of dominant electronic configurations remains unexplored, this factor is an important avenue for future investigation. The nuanced performance of $GW$ for multireference chemical systems is the primary finding of our analysis. Overall, $GW$ has to be used with care and there is room for improvement to make Green's function-based method accurate in systems with a strong multireference character. In this regard, the development of explicit multireference implementations, \cite{Brouder_2009,Linner_2019} although less black-box than the single-reference version, would be quite useful in certain chemical scenarios.

%%%%%%%%%%%%%%%%%%%%%%%%%%%%%%%%%
\section*{Supplementary Material}
%%%%%%%%%%%%%%%%%%%%%%%%%%%%%%%%%
See the \SupMat for the specification of the basis set and the total energies for the \ce{Be + H2} reaction, the total energies and additional $G_0W_0$ results of the set of multireference molecules, the geometry of the \ce{H6} cluster, and the IP of the \ce{HF} molecule as a function of the bond length.

%%%%%%%%%%%%%%%%%%%%%%%%%%
\section*{Acknowledgments}
%%%%%%%%%%%%%%%%%%%%%%%%%%
This project has received financial support from the European Research Council (ERC) under the European Union's Horizon 2020 research and innovation programme (Grant agreement no.~863481). Additionally, it was supported by the European Centre of Excellence in Exascale Computing (TREX), and has received funding from the European Union's Horizon 2020 --- Research and Innovation program --- under grant agreement no.~952165. HGAB was supported by Downing College, Cambridge, through the Kim and Julianna Silverman Research Fellowship.

%%%%%%%%%%%%%%%%%%%%%%%%%%%%%%%%%%%%%%
\section*{Data availability statement}
%%%%%%%%%%%%%%%%%%%%%%%%%%%%%%%%%%%%%%
The data that supports the findings of this study are available within the article and its supplementary material.

%%%%%%%%%%%%%%%%%%%%%
\section*{References}
%%%%%%%%%%%%%%%%%%%%%

%%%%%%%%%%%%%%%%%%%%
\bibliography{GW4MR}
%%%%%%%%%%%%%%%%%%%%

\end{document}